\documentclass{article}
\usepackage{graphicx}
\newcommand{\D}{d}
\newcommand{\E}{e}
\newcommand{\I}{i}
\renewcommand{\varrho}{\rho}
\renewcommand{\vec}[1]{{\mathbf #1}}

\begin{document}
\title{Random-matrix ensembles in $p$-wave vortices}
\author{Dmitri A. Ivanov\\
\it Institut f\"ur Theoretische Physik, \\
\it ETH-H\"onggerberg, CH-8093 Z\"urich, Switzerland}
\date{}

\maketitle              

\begin{abstract}
In disordered vortices in $p$-wave superconductors the
two new random-matrix ensembles may be realized: $B$
and $D$III-odd (of $so(2N+1)$ and $so(4N+2)/u(2N+1)$
matrices respectively). We predict these ensembles from
an explicit analysis of the symmetries of 
Bogoliubov--deGennes equations in three examples of
vortices with different $p$-wave order parameters.
A characteristic feature of the novel symmetry classes
is a quasiparticle level at zero energy. Class $B$ is
realized when the time-reversal symmetry is broken, and
class $D$III-odd when the time-reversal symmetry is preserved.
We also suggest that the main contribution to disordering
the vortex spectrum comes from the distortion of
the order parameter around impurities.
\end{abstract}

\bigskip

Since Wigner's modeling Hamiltonians of complex nuclei by
random matrices~\cite{Wigner}, the random-matrix theory
(RMT) has played an important role in studying mesoscopic
systems. In many cases, a chaotic (non-integrable) mesoscopic
system may be accurately described by RMT. The supersymmetric
technique by Efetov~\cite{Efetov} provides a microscopic
explanation of the RMT approximation for disordered systems.

According to the RMT approximation, the only characteristics
of the system affecting the eigenvalue correlations at small
energy scales are its symmetries. Therefore, a problem
arises of classifying symmetries of random-matrix ensembles.
It has been suggested by several authors~\cite{Altland,Zirnbauer-2,Caselle}
that random-matrix ensembles may be classified as corresponding
symmetric spaces. The symmetric spaces may be
divided into twelve infinite series reviewed in Table~1
(we split class $D$III in two subclasses: $D$III-even and
$D$III-odd)~\cite{Helgason}. Each of the
symmetry classes may occur in one of the three forms:
positive-curvature, negative-curvature and flat~\cite{Caselle}. 
The corresponding
Jacobians in the matrix space are expressed in terms
of trigonometric, hyperbolic, and polynomial functions
respectively. In the present paper we shall only discuss
the RMT for Hamiltonians forming a linear space and
therefore described by zero-curvature (flat) versions
of RMT.

\begin{table}[t]
\caption{Symmetric spaces and universality classes of random-matrix
ensembles: 
(\textbf{a})~the 
three Wigner-Dyson classes $A$ (unitary), $A$I (orthogonal), 
$A$II (symplectic)~\cite{Dyson};
(\textbf{b})~chiral 
classes $A$III (unitary), $BD$I (orthogonal), $C$II (symplectic),
with dimensions $p\le q$ have $q-p$ zero modes~\cite{Verbaarschot};
(\textbf{c})~superconducting 
classes $C$, $D$, $C$I, $D$III-even~\cite{Altland};
(\textbf{d})~$p$-wave vortex 
classes $D$III-odd, $B$ have one zero mode
}
\begin{center}
\renewcommand{\arraystretch}{1.4}
\setlength\tabcolsep{4pt}
\begin{tabular}{cccccc}
\hline\noalign{\smallskip}
\renewcommand{\arraystretch}{1}
\begin{tabular}{c}
  Cartan \\ class \\
\end{tabular} & 
\renewcommand{\arraystretch}{1}
\begin{tabular}{c}
  Symmetric \\ space \\
\end{tabular} &
Dimension &
Rank &
\hspace*{5pt}$\beta$\hspace*{5pt}
& $\alpha$ 
\\ [10pt]
\hline
\noalign{\smallskip}
$A$ [GUE]     & SU($N$) & $N^2{-}1$ & $N{-}1$ & 2 & $-$ \\
$A$I [GOE]    & SU($N$)/SO($N$) & $(N{-}1)(N{+}2)/2$ 
              & $N{-}1$ & 1 & $-$ \\
$A$II [GSE]   & SU($2N$)/Sp($N$) & $(N{-}1)(2N{+}1)$ 
              & $N{-}1$ & 4 & $-$ \\
$A$III [chGUE] & SU($p$+$q$)/S(U($p$)${\times}$U($q$)) & 
              $2pq$ & $p$ & 2 & $1{+}2(q{-}p)$ \\
$BD$I [chGOE] & SO($p$+$q$)/SO($p$)${\times}$SO($q$) & 
              $pq$ & $p$ & 1 & $q{-}p$ \\
$C$II [chGSE] & Sp($p$+$q$)/Sp($p$)${\times}$Sp($q$) & 
              $4pq$ & $p$ & 4 & $3{+}4(q{-}p)$ \\
              $C$ & Sp($N$) & $N(2N{+}1)$ & $N$ & 2 & 2 \\
$D$           & SO($2N$) & $N(2N{-}1)$ & $N$ & 2 & 0 \\
$C$I          & Sp($N$)/U($N$) & $N(N{+}1)$ & $N$ & 1 & 1 \\
$D$III-even   & SO($4N$)/U($2N$) & $2N(2N{-}1)$ & $N$ & 4 & 1 \\
$D$III-odd    & SO($4N$+2)/U($2N$+1) & $2N(2N{+}1)$ & $N$ & 4 & 5 \\
$B$           & SO($2N$+1) & $N(2N{+}1)$ & $N$ & 2 & 2 \\
\hline
\end{tabular}
\end{center}
\label{table1}
\end{table}

The simplest examples of the RMT symmetry classes are the three
Wigner-Dyson classes: unitary, orthogonal and symplectic
($A$, $A$I, and $A$II, respectively, in Cartan's notation)~\cite{Dyson}. 
In these classes, the energy level correlations
are invariant under translations in energies (at energy scales much
smaller than the spectrum width), and the
joint probability distribution of the energy levels $\omega_i$
is 
\begin{equation}
{\D P\{\omega_i\} \over \prod_i \D\omega_i}
\propto
\prod_{i<j} |\omega_i-\omega_j|^\beta \; .
\end{equation}
The parameter $\beta$ determines the strength of level
repulsion and takes values 2, 1, and 4 for the unitary,
orthogonal, and symplectic ensembles, respectively.

Other symmetry classes appear when there exists an additional
symmetry relating energies $E$ and $-E$. In the corresponding
random-matrix ensembles, the levels $\omega_i$ repel not only
each other, but also their mirror images $-\omega_i$, and the
Jacobian has the form
\begin{equation}
{\D P\{\omega_i\} \over \prod_i \D \omega_i}
\propto
\prod_{i<j} |\omega_i^2-\omega_j^2|^\beta
\prod_i \omega_i^\alpha \; .
\label{probability-alpha-beta}
\end{equation}
It is now characterized by the two parameters $\beta$ and $\alpha$
(the latter responsible for suppressing the density of states
near $E=0$).

The three chiral classes ($A$III, $BD$I, and
$C$II) appear in systems with a chiral symmetry anticommuting with the
Hamiltonian~\cite{Verbaarschot}. Four more classes
($C$, $D$, $C$I, and $D$III) describe mesoscopic superconducting
systems (with or without spin-rotational and 
time-reversal symmetries)~\cite{Altland}. 
The present paper is devoted to the last
two lines in Table~1: classes $D$III-odd and $B$. We shall
demonstrate that these two classes correspond to the symmetries
of a $p$-wave vortex with or without time-reversal symmetry,
respectively (the symmetry class $B$ in $p$-wave vortices has
also been predicted in~\cite{Bocquet}; a partial account of the
present work appeared as a preprint~\cite{Ivanov-energylevels}).

Below we consider three particular $p$-wave vortices and describe
their symmetries: the single-quantum vortex in the $A$-phase
with a fixed spin orientation of the order parameter, 
the half-quantum vortex in the $A$-phase with rotating orientation of
the order parameter, and the spin (fluxless) vortex  
in the $B$-phase. We explicitly
demonstrate that in the first two examples the symmetries of the
Hamiltonian are of type $B$, and in the last example -- of type
$D$III-odd. 
Then we briefly discuss microscopic requirements for the RMT
limit and show that the main contribution to the level mixing 
comes from inhomogeneous suppression of the order parameter by
impurities. Our estimates suggest that the RMT limit may be
achieved in a moderately clean superconductor.

\section{Single-Quantum Vortex}

In this section we consider a vortex in the order parameter 
similar to the $A$ phase of $^3$He. Namely, we assume that the
condensate wave function has the form
\begin{equation}
\Psi_\pm = \E^{\I \varphi}
\bigg[ 
d_x \Big( \left|\uparrow\uparrow\right\rangle + 
\left|\downarrow\downarrow\right\rangle \Big) +
\I d_y \Big( \left|\uparrow\uparrow\right\rangle - 
\left|\downarrow\downarrow\right\rangle \Big) +
d_z \Big( \left|\uparrow\downarrow\right\rangle + 
\left|\downarrow\uparrow\right\rangle \Big) 
\bigg]
(k_x \pm \I k_y) \; ,
\label{A-phase}
\end{equation}
and that the vector $\vec{d}$ defining the orientation of the
triplet is {\it fixed} (without loss of generality, take it to be
in the $\vec{z}$ direction. We also assume a fixed chirality of
the order parameter and take positive sign in place of $\pm$
in (\ref{A-phase}). Then the only allowed vortices are those
involving a rotation of the phase of the order parameter, and the magnetic 
flux in such vortices is quantized in conventional superconducting flux
quanta $\Phi_0=hc/2e$.

Below we analyze the low-lying energy levels
obtained as solutions to Bogo\-liubov--deGennes equations for
the mean-field Hamiltonian
\begin{equation}
H=\sum_\alpha \Psi^\dagger_\alpha \left[{(\vec{p} -e \vec{A})^2 \over 2m}
+V(\vec{r})-\varepsilon_F\right] \Psi_\alpha
+ \Psi_\uparrow^\dagger \left( \Delta_x * {p_x\over k_F} +
\Delta_y * {p_y\over k_F} \right)
\Psi_\downarrow^\dagger + {\rm h.c.} \; ,
\label{Hamiltonian}
\end{equation}
where $\Psi_\alpha$ are the electron operators ($\alpha$ is the spin
index), $V(\vec{r})$ is the external potential of impurities, 
$\vec{A}(\vec{r})$ is
the electromagnetic vector potential, $\Delta_x(\vec{r})$ and 
$\Delta_y(\vec{r})$ are
the coordinate-dependent components of the superconducting gap.
[In the bulk, the preferred superconducting order is one of the two
chiral components $\eta_\pm=\Delta_x\pm\I\Delta_y$, but in
inhomogeneous systems, such as a vortex core, an admixture
of the opposite component may be self-consistently 
generated~\cite{Heeb}. We account for this effect by allowing the
two independent order parameters $\Delta_x$ and $\Delta_y$.]
Star ($*$) denotes the symmetrized ordering of the gradients $p_\mu$ and the
order parameters $\Delta_\mu$ [definition: $A*B\equiv (AB+BA)/2$]. 
At infinity, the order parameter
imposes the vortex boundary conditions:
\begin{equation}
\Delta_x(r\to\infty,\phi)=\Delta_0 \E^{\pm \I\phi}, \qquad
\Delta_y(r\to\infty,\phi)=\I\Delta_0 \E^{\pm \I\phi} \; ,
\label{boundary-conditions}
\end{equation}
where $r$ and $\phi$ are polar coordinates.
Plus or minus signs in the exponent correspond to a positive
or a negative single-quantum vortex.
For an axially-symmetric vortex with the chirality of the
order parameter non-self-consistently fixed ($\Delta_y \equiv \I \Delta_x$), 
without the vector-potential $\vec{A}(\vec{r})$ and without disorder 
$V(\vec{r})$,
the low-lying eigenstates of the Hamiltonian (\ref{Hamiltonian})
have been found by Kopnin and Salomaa~\cite{Kopnin}. 
The spectrum is 
\begin{equation}
E_n=n \omega_0 \; , \quad 
(p-{\rm wave}) \; , \qquad 
n=0,\pm 1, \pm 2, \dots
\label{p-spectrum}
\end{equation}
with $\omega_0 \sim \Delta^2/\varepsilon_F$.
This result should be compared with the spectrum of the
vortex core in a $s$-wave superconductor~\cite{deGennes}: 
\begin{equation}
E_n=\left(n+{1\over 2}\right) \omega_0 \; , 
\quad (s-{\rm wave})\; ,
\qquad n=0,\pm 1, \pm 2, \dots
\label{s-spectrum}
\end{equation}

The common feature of the spectra in the $s$-wave and $p$-wave cases
is the symmetry with respect to zero energy. If we interpret holes in the
negative-energy levels as excitations with positive energies
(and with opposite spin), then this symmetry implies that
the excitations are doubly degenerate in spin: to each spin-up
excitation there corresponds a spin-down excitation at the same energy.
For the $s$-wave vortex, this degeneracy is due to the full spin-rotation
$SU(2)$ symmetry. The $p$-wave Hamiltonian (\ref{Hamiltonian})
has a reduced spin symmetry. Namely, it has the symmetry group
$O(2)$ generated by rotations about the $z$-axis 
($\Psi_\uparrow \mapsto \E^{\I\alpha}\Psi_\uparrow$, 
$\Psi_\downarrow \mapsto \E^{-\I\alpha}\Psi_\downarrow$) and by the
spin flip $\Psi_\uparrow \mapsto \Psi_\downarrow$, $\Psi_\downarrow
\mapsto \Psi_\uparrow$. This non-abelian group causes the
two-fold degeneracy of all levels, except for the zero-energy
level(s) where the symmetry $O(2)$ may mix the creation
and annihilation operators for the same state. This symmetry
is crucial for our discussion. 
Note that we have not included in the Hamiltonian neither the spin-orbit
term $(\vec{U}_\mathrm{SO} \cdot [\vec{\sigma} \times \vec{p}])$,
nor the Zeeman splitting $\vec{H}(\vec{r}) \cdot \vec{\sigma}$. 
Either of these terms would break the spin
symmetry $O(2)$, which would eventually result in a different universality
class of the disordered system (type $D$ with non-degenerate levels),
if these terms are sufficiently strong.

The difference between the $s$- and $p$-wave vortices is the zero-energy 
level in the $p$-wave case. It has been shown by Volovik that this
level has a topological nature~\cite{Volovik-zeromodes}. Indeed, suppose we
gradually increase disorder in the Hamiltonian (\ref{Hamiltonian}).
The levels shift and mix, but the degeneracy of the levels remains
the same as long as the symmetry $O(2)$ is preserved. The total
number of levels remains {\it odd}, and therefore the zero-energy
level {\it cannot shift} if the final Hamiltonian is a continuous
deformation of the original one (without disorder), i.e. if the
topological class of the boundary conditions (\ref{boundary-conditions})
remains the same.

Now we can identify the symmetry class of the Bogoliubov--deGennes
Hamiltonian.
The only symmetry of the mean-field Hamiltonian (\ref{Hamiltonian}) 
is the spin symmetry $O(2)$. 
The time-reversal symmetry is already broken by the vortex
and by the pairing, and therefore neither
the vector potential $\vec{A}(x)$ nor local deformations of $\Delta_\mu$
can reduce the symmetry of the Hamiltonian. When projected onto
spin-up excitations $\gamma^\dagger_\uparrow = 
\int[u(\vec{r})\Psi^\dagger_\uparrow(\vec{r}) 
+ v(\vec{r})\Psi_\downarrow(\vec{r})] d^2 \vec{r}$,
the Bogoliubov--deGennes Hamiltonian 
for the two-component vector $(u,v)$ takes the form:
\begin{equation}
H_\mathrm{BdG}=\pmatrix{
\left[ {(-\I\nabla - e \vec{A})^2 \over 2m} + V(\vec{r}) 
-\varepsilon_F \right] &
\left[{\Delta_x\over k_F}*(-\I\nabla_x)+
{\Delta_y\over k_F}*(-\I\nabla_y)\right]
\cr
\left[{\Delta^*_x\over k_F}*(-\I\nabla_x)+
{\Delta^*_y\over k_F}*(-\I\nabla_y)\right] &
- \left[ {(-\I\nabla + e \vec{A})^2 \over 2m} + V(\vec{r}) 
-\varepsilon_F \right] }\; .
\label{BdGequation}
\end{equation}
In an arbitrary orthonormal basis of electronic states, this Hamiltonian
may be written as a matrix
\begin{equation}
H_\mathrm{BdG}=\pmatrix{
h & \Delta \cr
\Delta^\dagger & -h^*}\; .
\label{Ham-short}
\end{equation}
From the hermiticity of the Hamiltonian, it follows that $h^\dagger=h$.
From the explicit form of the $p$-wave pairing, $\Delta = -\Delta^T$
(it is here that the $p$-wave structure of the pairing is important;
for $s$-wave pairing we would have $\Delta = \Delta^T$ instead).
These are the only restrictions on the Hamiltonian (\ref{Ham-short}).
If we define 
\begin{equation}
U_0={1\over\sqrt2}\pmatrix{1&1\cr \I &-\I},
\end{equation}
the restrictions on the Hamiltonian (\ref{Ham-short})
are equivalent to the condition that the rotated matrix
$\I U_0 H_\mathrm{BdG} U_0^{-1}$ is real antisymmetric, i.e. it belongs to the
Lie algebra $so(M)$, where $M$ is the dimension of the 
Hilbert space (the same rotation of the Hamiltonian was used
in~\cite{Altland} to identify the $D$ universality class).

The last step in our argument is to note that, under the vortex boundary
conditions, the dimension of the Hamiltonian (\ref{Ham-short})
is odd, not even (this may be difficult to visualize from the
particle-hole representation (\ref{Ham-short}), but easier
from the rotated Hamiltonian $U_0 H_\mathrm{BdG} U_0^{-1}$). 
Thus for a single-quantum
vortex, we identify the space of the Hamiltonians as $so(2N+1)$
(class $B$ in Cartan's notation, see Table~1).

\section {Half-Quantum Vortex}

\begin{figure}[t]
\begin{center}
\includegraphics[width=0.75\textwidth]{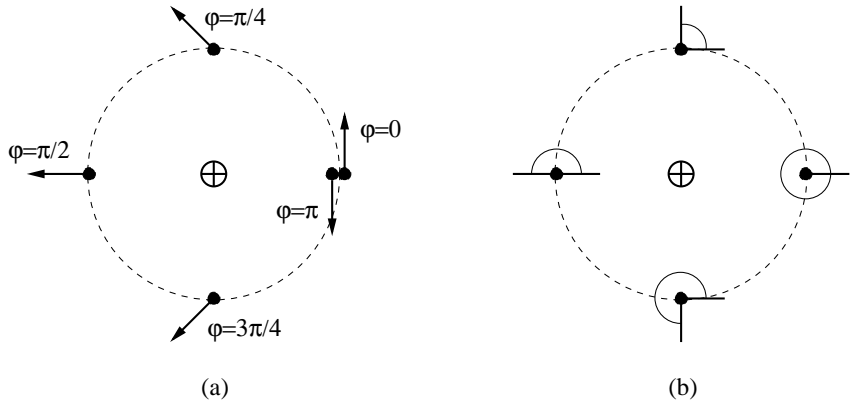}
\end{center}
\caption[]{(\textbf{a}) Half-quantum vortex. 
Arrows denote the direction of vector $\vec{d}$. 
(\textbf{b}) Spin vortex. The angle $\theta$ is shown}
\label{fig1}
\end{figure}

Now let us consider a slightly different order parameter: 
suppose that the condensate wave function is described by
the same expression (\ref{A-phase}), but now the vector
$\vec{d}$ is able to rotate (either in plane or in all three
dimensions). The order parameter does not change under 
simultaneous change of sign of the vector $\vec{d}$ and 
phase shift by $\pi$ of the phase $\varphi$: $(\varphi, \vec{d})
\sim (\varphi+\pi, -\vec{d})$. This makes possible a half-quantum
vortex with magnetic flux $\Phi_0/2$~\cite{Volovik-Mineev}. 
On going around such a vortex,
both the vector $\vec{d}$ and the phase $\varphi$ rotate by
$\pi$ (Fig.~\ref{fig1}a) (see also discussion of such vortices 
in~\cite{Volovik-halfquantum,Ivanov-nonabelian}).

Consider first the case when the vector $\vec{d}$ rotates in 
a plane (without loss of generality, let $\vec{d}$ rotate in the
$x$--$y$ plane). Then the condensate wave function (\ref{A-phase})
is the sum of two decoupled components containing 
$\left|\uparrow\uparrow\right\rangle$ and 
$\left|\downarrow\downarrow\right\rangle$ pairings:
\begin{equation}
\Psi(r,\theta) = 
\left[ \Delta_+^{(1)}(r,\theta) \left|\uparrow\uparrow\right\rangle + 
\Delta_+^{(2)}(r,\theta) \left|\downarrow\downarrow\right\rangle \right]
(k_x + \I k_y)
\end{equation}
[here $r$ and $\theta$ are the polar coordinates].
In a half-quantum vortex, one of the pairing components 
$\Delta_+^{(1)}$ and $\Delta_+^{(2)}$ has an odd winding number
of the phase around the vortex, while the other has an
even winding number of the phase. In the simplest vortex
of this type, the even winding number is zero, the corresponding
component of the order parameter is nearly uniform and
does not produce states near Fermi energy. The component
with an odd winding number (without loss of generality,
let it be $\Delta_+^{(1)}$, and the winding number be one)
has vortex boundary conditions. The mean-field Hamiltonian
decouples and the vortex part of the Hamiltonian (spin-up sector)
takes the form [we have also assumed an axially symmetric order
parameter]:
\begin{equation}
H=\int \D^2\vec{r} \left[
\Psi_\uparrow^\dagger \left({\vec{p}^2\over 2m} -\varepsilon_F\right)
\Psi_\uparrow + \E^{\I\theta} \Delta(r) \Psi_\uparrow^\dagger
\left(\nabla_x+\I\nabla_y\right)\Psi_\uparrow^\dagger + {\rm h.c.} 
\right]\; .
\label{ham-halfquantum}
\end{equation}
In this form the half-quantum vortex is equivalent to a single-quantum
vortex in a superconductor of spinless fermions (such a superconductor
must have odd pairing). The quasiparticles do not have a definite spin
projection, but mix particles and holes in the same spin-up sector:
$\gamma^\dagger = 
\int[u(\vec{r})\Psi^\dagger_\uparrow(\vec{r}) 
+ v(\vec{r})\Psi_\uparrow(\vec{r})] \D^2 \vec{r}$.
The Bogoliubov--deGennes equations for $u$ and $v$ 
turn out to be identical to those for a single-quantum
vortex (\ref{BdGequation}), and the solution of 
Kopnin and Salomaa~\cite{Kopnin} is equally applicable
to this half-quantum vortex resulting in the same spectrum
(\ref{p-spectrum}). However, because of the relation
$\gamma^\dagger(E)=\gamma(-E)$, the number of subgap states
in the half-quantum vortex is one half of those 
in a single-quantum vortex: particles in negative-energy levels are
identical to holes in positive-energy levels. 
The zero-energy state is a Majorana 
fermion~\cite{Ivanov-nonabelian,Read-Green}.

Now add a disorder term to the Hamiltonian of a clean vortex
(\ref{ham-halfquantum}). Consider the
following perturbations:
\begin{itemize}
\item 
potential scattering: \qquad 
$H_V = \Psi^\dagger_\alpha V(\vec{r}) \Psi_\alpha$,
\item
Zeeman field: \qquad 
$H_{\vec{h}} = \Psi^\dagger_\alpha  
\left( \vec{h}(\vec{r}) \vec{\sigma}_{\alpha\beta} \right) 
\Psi_\beta$,
\item
vector potential: \qquad 
$H_{\vec{a}} = \Psi^\dagger_\alpha 
\left( \vec{a}(\vec{r}) * \vec{p} \right) 
\Psi_\alpha$,
\item
spin-orbit scattering: \qquad 
$H_{\vec{U}} = \Psi^\dagger_\alpha 
\left( \vec{U}_\mathrm{SO} (\vec{r}) * [\vec{p} \times 
\vec{\sigma}_{\alpha\beta}] \right) 
\Psi_\beta$,
\item
deformation of order parameter: \\ \hspace*{2cm} 
$H_\Delta = \Psi^\dagger_\alpha 
\left( \Delta^{(x)}_{\alpha\beta}(\vec{r}) * p_x +
 \Delta^{(y)}_{\alpha\beta}(\vec{r}) * p_y \right)
\Psi^\dagger_\beta + {\rm h.c.}$
\end{itemize}
We assume that the perturbation may be switched on adiabatically
in such a way that subgap levels stay localized. One
possible way to fulfill this condition is to require that
the perturbation is introduced only in a finite region
around the vortex core.

The symmetry classification proceeds slightly 
differently depending on whether
the perturbation preserves the decoupling of the Hamiltonian
into spin-up and spin-down sectors.

The Hamiltonian remains decoupled under the perturbations
$H_V$, $H_{\vec{a}}$, $H_{\vec{h}}$ with $\vec{h} \parallel \hat{\vec{z}}$,
 $H_{\vec{U}}$ with $\vec{U}_\mathrm{SO} \perp \hat{\vec{z}}$, and
$H_\Delta$ with diagonal $\Delta_{\alpha\beta}$. With these
perturbations, the ``spin-up'' sector of the Hamiltonian in
the basis $(u,v)$ preserves its form (\ref{Ham-short}) with
the same symmetries as for a single-flux vortex. The symmetry
class of the Hamiltonian is therefore identified as
$so(2N+1)$ (class $B$). 

If other perturbations are present, so that the Hamiltonian
no longer splits into ``spin-up'' and ``spin-down'' parts,
the full matrix of the Hamiltonian needs to be considered.
Without any spin structure, the Hamiltonian breaking time-reversal
and spin symmetries belongs to the $so(M)$ symmetry class,
as shown by Altland and Zirnbauer~\cite{Altland}. Their argument
also works in our case, with the reservation that the number of 
levels $M$ is odd (with each level counted twice: once as a particle
and the other time as a hole). It follows from the usual continuity
consideration (the Majorana fermion at zero energy survives any
local perturbation). Therefore the symmetry class is again 
$so(2N+1)$ (class $B$). [In the examples considered by Altland and
Zirnbauer, the number of levels was even, which lead to the symmetry
class $D$ or $so(2N)$].

It is important however, that to reach the RMT limit in the latter
case, such a system must have sufficiently strong disorder to bring
the quasiparticles from the ``spin-down'' sector (gapped in the
clean vortex) to the Fermi energy and to mix them strongly with
the quasiparticles in the ``spin-up'' sector. In the present paper,
we do not discuss conditions for reaching this limit.

\section{Spin Vortex}

In this section we consider a hypothetical phase of a triplet
superconductor analogous to the phase $B$ of $^3$He. Namely, the
condensate wave function is assumed to be
\begin{equation}
\Psi= \Delta \E^{\I\varphi}
\bigg[ 
d_x(\vec{k}) \Big( \left|\uparrow\uparrow\right\rangle + 
\left|\downarrow\downarrow\right\rangle \Big) +
\I d_y(\vec{k}) \Big( \left|\uparrow\uparrow\right\rangle - 
\left|\downarrow\downarrow\right\rangle \Big) 
\bigg]\; , \qquad \vec{d}(\vec{k})=R_{\theta}(\vec{k})\; ,
\label{B-phase}
\end{equation}
and $R_\theta$ is a rotation in the plane by angle $\theta$
(like in the previous sections, we consider a two-dimensional
problem, and $\vec{k}$ is a two-dimensional vector). 
In this phase, the vector $\vec{d}$
rotates a full turn as the vector $\vec{k}$ goes around
the circular Fermi surface. The parameter $\theta$ denotes the
angle between vectors $\vec{k}$ and $\vec{d}$. We
neglect spin-orbit interactions and assume that all values
of $\theta$ are allowed.

It is important that this phase preserves the time-reversal symmetry.
Moreover, a vortex structure is possible without time-reversal
symmetry breaking. It is realized by rotating the angle $\theta$
a full turn of $2\pi$ when going around the vortex center 
and keeping the phase $\varphi$ constant (Fig.~\ref{fig1}b). 
This vortex preserves the time-reversal symmetry and is
a topological defect in the spin structure of the triplet
pairing, thus we shall call it ``spin vortex''.

The Hamiltonian of the clean axially symmetric spin vortex
splits into ``spin-up'' and ``spin-down'' sectors, similarly
to the case of the half-quantum vortex considered in the
previous section. In the spin vortex, both the ``spin-up''
and the ``spin-down'' components have vortex structure.
The ``spin-up'' Hamiltonian is identical to the
Hamiltonian $H$ of (\ref{ham-halfquantum}), while
the ``spin-down'' component is its complex conjugate $H^*$
(with spin reversed) --- or vice versa.

In a clean spin vortex, the spectrum of each of the two vortices
(in the ``spin-up'' and the ``spin down'' sectors) are identical
to that in the half-quantum vortex of the previous section,
and therefore the total spectrum coincides with that of the single-flux
vortex of Section~I (with the two Majorana fermions combining into a single 
zero-energy level). In contrast to the single-quantum vortex,
it is now not the spin symmetry which is responsible for the
double degeneracy of the levels with non-zero energy, but the
time-reversal symmetry (the degeneracy is due to Kramers'
theorem).

Now suppose that the Hamiltonian is perturbed by a disorder term
preserving the time-reversal symmetry. The allowed perturbations include
potential scattering $H_V$, spin-orbit scattering $H_{\vec{U}}$, 
as well as deformation of the order parameter $H_\Delta$
in a time-reversal invariant  way. Then the levels with non-zero
energy stay doubly degenerate, and the zero-energy level cannot
shift (because it is non-degenerate). The symmetry class of the
perturbed Hamiltonian may be identified from the argument
of Altland and Zirnbauer as $D$III (time-reversal symmetry present,
spin symmetry broken)~\cite{Altland}. In the $p$-wave vortex, the number
of subgap levels is odd, and therefore the symmetry class is
$D$III-odd, in contrast to class $D$III-even studied by
Altland and Zirnbauer [similarly to distinguishing between classes
$B$ and $D$, we find it useful to distinguish between
classes $D$III-even and $D$III-odd: they have different parameters 
$\alpha$, as a consequence of the zero-energy level in odd dimensions].

\section{Level Mixing by Disorder}

In this section we discuss whether the RMT limit may be realized
in disordered vortices. We suppose that the superconductor
contains a finite concentration $n_\mathrm{imp}$ 
of spinless impurities with strong
electron scattering (Born parameter is of order one).
Then the effect of impurities is dual:
they directly contribute a potential term $H_V$ in the
Hamiltonian and they also suppress the superconducting
order parameter thus inducing an inhomogeneous pairing
perturbation $H_\Delta$. We shall see that the latter
effect has a much stronger influence on shifting the
energy levels of the subgap states.

Consider first the potential term $H_V$ with the impurity
potential $V(\vec{r})=\sum_i V_i \delta(\vec{r}-\vec{r}_i)$.
The matrix element of the impurity potential between the two localized
states is equal to 
\begin{equation}
(H_V)_{mn}=\sum_i V_i \varrho_{mn}(\vec{r}_i)\; ,
\qquad \varrho_{mn}(\vec{r})=\left(u^*_m(\vec{r}) u_n(\vec{r}) - 
v^*_m(\vec{r}) v_n(\vec{r})\right)\; .
\end{equation}
From explicitly solving Bogoliubov--deGennes equations for a clean
$p$-wave vortex~\cite{Kopnin}, one finds that $u_n$ and $v_n$
are proportional to each other, to the leading order in $E/\Delta$
(where $E=n\omega_0$ is the energy of the $n$-th level).
Therefore the matrix element $\varrho_{mn}$ between any two
low-lying states nearly vanishes (this was noticed by Volovik
in~\cite{Volovik-zeromodes}). A more accurate calculation shows
that matrix elements of the charge density between states at energy
$E$ is of the order of $\xi^{-2}(E/\Delta)$, which is by the factor
$(E/\Delta)$ smaller than that in the $s$-wave vortex.
As a consequence, even for strong impurities (with 
$V_i\sim \varepsilon_F k_F^{-2}$), an extremely high impurity
concentration ($n_\mathrm{imp}\sim k_F^2$) is required to mix
low-energy levels --- a much higher disorder than that destroying
$p$-wave superconductivity 
($n_\mathrm{imp}\sim k_F\xi^{-1}$)~\cite{Balian,Ueda}.
Therefore, for a uniformly disordered $p$-wave superconductor, 
the $H_V$ term is insufficient for the RMT regime in the vortex core.

The suppression of the order parameter $H_\Delta$ turns out to be
a more important effect. In unconventional superconductors, spinless
impurities suppress pairing in the region of size $\xi$
which is of the same order as the size of the vortex core.
For a homogeneous order parameter (without vortex), the suppression
of the order parameter $\delta\Delta(r)$ decreases as $R^{-1}$
(in our two-dimensional problem) at distances up to $\xi$, 
and is negligible at distances much larger than 
$\xi$~\cite{Rainer,Choi}, and the integral suppression may be estimated as
$\int\delta\Delta\; \D^2 \vec{r} \sim \Delta\xi k_F^{-1}$.
Also, in a chiral superconductor, the component of the
order parameter of opposite chirality is admixed in the vicinity of
impurity~\cite{Okuno}. The two effects gives comparable
contributions to the matrix elements between subgap states.
An accurate calculation of the impurity influence on the
states in the vortex core requires a full self-consistent solution
for the order parameter taking into account both the vortex
and the impurities. This goes beyond the scope of this paper,
and we only estimate the order of magnitude of the corresponding
matrix elements. The deformation  of the order parameter $H_\Delta$
has the matrix elements between subgap states $m$ and $n$:
\begin{eqnarray}
(H_\Delta)_{mn} = \int \D^2\vec{r} \,
\bigg( &&\!\! u_m^*(\vec{r}) \left[\delta\Delta^{(\alpha)}(\vec{r})
* (\I\nabla_\alpha)\right] u_n(\vec{r}) \nonumber\\
& &+ v_m^*(\vec{r}) \left[\delta{\Delta^{(\alpha)}}^*(\vec{r})
* (\I\nabla_\alpha)\right] v_n(\vec{r}) \bigg)\; .
\end{eqnarray}
Unlike the case of the matrix elements of $H_V$, there is no
cancellation in $(H_\Delta)_{mn}$, and we may estimate, for
one impurity in the vortex core
\begin{equation}
(H_\Delta)_{mn}\sim{1\over \xi^2} \int \delta\Delta(\vec{r})\, \D^2 \vec{r}
\sim \Delta (k_F\xi)^{-1} \sim \omega_0\; .
\end{equation}
Thus, a single impurity introduces matrix elements of the order
of interlevel spacing. 
These estimates suggest that the RMT regime may be achieved
when the number of impurities in the vortex core is much 
greater than one (which corresponds to the
moderately clean regime $\Delta^{-1}\ll\tau\ll\omega_0^{-1}$).

\section{Summary and Discussion}

We have shown that symmetries of vortex Hamiltonians in
$p$-wave superconductors correspond to one of the two 
random-matrix ensembles: $B$ or $D$III-odd. These two RMT
ensembles are distinguished from the rest of ensembles by
the presence of zero-energy modes (zero-energy modes 
are also present in chiral ensembles at $p\ne q$, see Table~\ref{table1}).
Zero-energy levels appear in $p$-wave vortices as a consequence
of odd pairing symmetry combined with the vortex (topologically
nontrivial) boundary conditions. We have checked the symmetries
of the Hamiltonian explicitly taking three vortices as examples:
a single-quantum vortex with spin symmetry, a half-quantum vortex,
and a flux-less (spin) vortex without time-reversal symmetry breaking.
Based on these examples, we may conjecture that classes $B$ and $D$III-odd
appear in any vortex-like structure with odd pairing, whenever
a zero-energy mode is present. In cases with time reversal symmetry,
the vortex belongs to the class $D$III-odd; when the time-reversal
symmetry is broken --- to the class $B$. Of course, this can also
be checked explicitly in any particular case.

The two classes $B$ and $D$III-odd considered in this paper are
odd-dimensional counterparts of the classes $D$ and $D$III-even,
respectively (see Table~1), with the even-dimensional classes
realized in conventional (singlet-pairing) 
normal-superconducting structures~\cite{Altland}.
Far from zero energy, the statistics of energy levels is not
affected by the zero mode. In the immediate vicinity of 
zero energy, repulsion from the zero mode in classes $B$ and $D$III-odd
increases the exponent $\alpha$ (see Table~1), suppressing the
average density of states $\langle \varrho(\omega) \rangle$ as
$\langle \varrho(\omega) \rangle \propto \omega^2$ for class $B$, and
$\langle \varrho(\omega) \rangle \propto |\omega|^5$ for class $D$III-odd.

Because of the zero modes, the classes $B$ and $D$III-odd stand
somewhat separately from the rest of RMT ensembles. In particular,
they do not appear in the table of correspondence between symmetries
of the Hamiltonian and those of the transfer matrix~\cite{Caselle,Mudry}
(for RMT ensembles without zero-energy levels, this table establishes
a one-to-one correspondence). Neither these two ensembles are known
to be derived from a supersymmetric sigma-model as other classes 
are~\cite{Zirnbauer-2}. Understanding the latter fact remains
an interesting problem, as well as a possible microscopic derivation
of the random-matrix theory for $p$-wave vortices with the
supersymmetric method, analogously to that for the $s$-wave 
vortex~\cite{Bundschuh,Skvortsov}.

Several more comments can be made on comparing our results to the
level statistics in $s$-wave vortices. The latter is known to
belong to the symmetry class $C$~\cite{Bundschuh,Skvortsov}
having spin-rotational symmetry and broken time-reversal 
symmetry~\cite{Altland}. In contrast with the $SU(2)$ spin symmetry
in conventional superconductors, the single-quantum vortex
considered in our paper has the $O(2)$ spin symmetry instead,
which leads to a somewhat different symmetry classification.
The resulting symmetry is $so(N)$, which leads to classes $D$ 
(in even dimensions, no vortex) or $B$ (odd dimensions, vortex
with zero mode).

The joint probability distributions of the energy levels 
(\ref{probability-alpha-beta}) for classes $B$ ($p$-wave vortex)
and $C$ ($s$-wave vortex) coincide (with $\alpha=\beta=2$),
and only the zero-energy level distinguishes the two ensembles.
In particular, the average density of states in the class $B$
random-matrix ensemble is 
\begin{equation}
\langle 
\varrho(\omega)\rangle ={1\over\omega_0} - {\sin(2\pi\omega / \omega_0) 
\over 2 \pi\omega}
+\delta(\omega).
\end{equation}
(for class $C$ it is the same but without the $\delta(\omega)$ term).

In $s$-wave vortices, Koulakov and Larkin found that 
in a wide range of impurity
concentrations, the level mixing does not lead to a class $C$,
but has more symmetries~\cite{Koulakov}. These symmetries arise
from the fact that the impurities are local scatterers in a 
chiral quasi-one-dimensional two-channel system, resulting in
an ensemble of $2\times 2$ random matrices, instead of
$N\times N$ matrices with large $N$. Such effect seems improbable
in $p$-wave vortices where the main contribution to level
mixing comes from the distortion of the order parameter around
impurities which is extended in space to distances of the order
of the core size. In this paper we performed only
an order-of-magnitude estimate suggesting that the RMT limit
may be reached in moderately clean vortices (with the number of
impurities per vortex core much greater than one). A
more accurate self-consistent treatment of impurities in the vortex core
is required for a quantitative study of the effects of disorder.


The author wishes to thank M.~V.~Feigel'man for 
suggesting this problem, discussions, and comments on the manuscript.
At different stages of work, the author benefited from discussions
with  G.~Blatter, V.~Geshkenbein, D.~Gorokhov, 
R.~Heeb, L.~Ioffe, C.~Mudry, 
M.~Skvortsov, G.~E.~Volovik, and M.~Zhitomirsky.
The author thanks Swiss National Foundation for financial support.



\begin{thebibliography}{27.}

\bibitem{Wigner}
E.~P.~Wigner: 
Proc.~Camb.~Phil.~Soc. \textbf{47}, 790 (1951)

\bibitem{Efetov}
K.~Efetov: \emph{Supersymmetry in disorder and chaos} (Cambridge University
Press, Cambridge 1997)

\bibitem{Altland}
A.~Altland, M.~R.~Zirnbauer:
Phys.~Rev. B \textbf{55}, 1142 (1997) [cond-mat/9602137]

\bibitem{Zirnbauer-2}
M.~R.~Zirnbauer:
J.~Math.~Phys. \textbf{37}, 4986 (1996)

\bibitem{Caselle}
M.~Caselle: A new classification scheme
for random matrix theories, cond-mat/9610017

\bibitem{Mudry}
P.~W.~Brouwer, C.~Mudry, A.~Furusaki:
Phys.~Rev.~Lett. \textbf{84}, 2913 (2000)
[cond-mat/9904200];\\
P.~W.~Brouwer, A.~Furusaki, I.~A.~Gruzberg, C.~Mudry:
Localization and delocalization in dirty 
superconducting wires, cond-mat/0002016

\bibitem{Helgason}
S.~Helgason: 
\emph{Differential geometry, Lie groups, and symmetric spaces}
(Academic Press, New York 1978)

\bibitem{Dyson}
F.~J.~Dyson:
J.~Math.~Phys. \textbf{3}, 1199 (1962)

\bibitem{Verbaarschot}
J.~Verbaarschot:
Phys.~Rev.~Lett. \textbf{72}, 2531 (1994)

\bibitem{Bocquet}
M.~Bocquet, D.~Serban, M.~R.~Zirnbauer:
Disordered 2d quasiparticles in class D: 
Dirac fermions with random mass, and dirty superconductors,
cond-mat/9910480

\bibitem{Ivanov-energylevels}
D.~A.~Ivanov: 
The energy-level statistics in the core of a vortex
in a $p$-wave superconductor, 
cond-mat/9911147

\bibitem{Heeb}
R.~Heeb, D.~F.~Agterberg:
Phys.~Rev.~B \textbf{59}, 7076 (1999) [cond-mat/9811190]

\bibitem{Kopnin}
N.~B.~Kopnin, M.~M.~Salomaa:
Phys.~Rev.~B \textbf{44}, 9667 (1991)

\bibitem{deGennes}
C.~Caroli, P.-G.~de~Gennes, J.~Matricon: 
Phys.~Lett. \textbf{9}, 307 (1964)

\bibitem{Volovik-zeromodes}
G.~E.~Volovik:
Pis'ma Zh.~Exp.~Teor.~Fiz. \textbf{70}, 601 (1999)
[JETP Lett. \textbf{70}, 609 (1999); cond-mat/9909426]

\bibitem{Volovik-Mineev}
G.~E.~Volovik, V.~P.~Mineev:
Pis'ma Zh.~Exp.~Teor.~Fiz. \textbf{24}, 605 (1976)
[JETP Lett. \textbf{24}, 561 (1976)]

\bibitem{Volovik-halfquantum}
G.~E.~Volovik:
Pis'ma Zh.~Exp.~Teor.~Fiz. \textbf{70}, 776 (1999)
[JETP Lett. \textbf{70}, 792 (1999); cond-mat/9911374].

\bibitem{Ivanov-nonabelian}
D.~A.~Ivanov: Non-abelian statistics of half-quantum vortices
in p-wave superconductors, cond-mat/0005069

\bibitem{Read-Green}
N.~Read, D.~Green:
Phys.~Rev. B \textbf{61}, 10267 (2000) [cond-mat/9906453]

\bibitem{Balian}
R.~Balian, N.~R.~Werthammer: 
Phys.~Rev. \textbf{131}, 1553 (1963)

\bibitem{Ueda}
K.~Ueda, T.~M.~Rice:
`Heavy electron superconductors ---
some consequences of the $p$-wave pairing'.
In: \emph{Theory of Heavy Fermions and Valence Fluctuations},
ed. by T.~Kasuya, T.~Saso (Springer, Berlin 1985)

\bibitem{Rainer}
D.~Rainer, M.~Vuorio: 
J.~Phys.~C: Solid State Phys. \textbf{10}, 3093 (1977)

\bibitem{Choi}
C.~H.~Choi, P.~Muzikar: 
Phys.~Rev. B \textbf{39}, 9664 (1989)

\bibitem{Okuno}
Y.~Okuno, M.~Matsumoto, M.~Sigrist: 
Analysis of
impurity-induced circular currents for the chiral superconductor,
cond-mat/9906093

\bibitem{Bundschuh}
R.~Bundschuh, C.~Cassanello, D.~Serban, M.~R.~Zirnbauer:
Nucl.~Phys. B \textbf{532}, 689 (1998) [cond-mat/9806172]

\bibitem{Skvortsov}
M.~A.~Skvortsov, M.~V.~Feigel'man, V.~E.~Kravtsov:
Pis'ma Zh.~Exp.~Teor.~Fiz. \textbf{68}, 78 (1998)
[JETP Lett. \textbf{68}, 84 (1998)]

\bibitem{Koulakov}
A.~A.~Koulakov, A.~I.~Larkin:
Phys.~Rev. B \textbf{60}, 14597 (1999) [cond-mat/9810125]

\end{thebibliography}
\end{document}